\documentclass[prl,superscriptaddress,a4paper,twocolumn]{revtex4-2}
\usepackage{geometry}
\usepackage[english]{babel}
\usepackage{hyperref}
\usepackage{amsfonts}
\usepackage{amsmath, amssymb, amsfonts, mathrsfs}
\usepackage{graphicx}
\usepackage{xspace}
\usepackage{xcolor}
\usepackage{multirow}
\usepackage{array} 
\usepackage{diagbox}

\usepackage{float}

\newcolumntype{P}[1]{>{\centering\arraybackslash}p{#1}}
\newcolumntype{M}[1]{>{\centering\arraybackslash}m{#1}}

\usepackage[bottom]{footmisc}
\newcommand{\ket}[1]{|#1\rangle}
\newcommand{\mean}[1]{\langle #1 \rangle}
\newcommand{\bra}[1]{ \langle #1 \,|}

\makeatletter
\renewcommand{\fnum@figure}{Fig. \thefigure}
\makeatother

\begin{document}


\title{Experimental demonstration of genuine tripartite nonlocality \\under strict locality conditions}

\author{Liang Huang$^{\ast}$}
\author{Xue-Mei Gu$^{\ast}$}
\affiliation{Hefei National Research Center for Physical Sciences at the Microscale and School of Physical Sciences, University of Science and Technology of China, Hefei 230026, China}
\affiliation{Shanghai Research Center for Quantum Science and CAS Center for Excellence in Quantum Information and Quantum Physics, University of Science and Technology of China, Shanghai 201315, China}
\affiliation{Hefei National Laboratory, University of Science and Technology of China, Hefei 230088, China}

\author{Yang-Fan Jiang$^{\ast}$}
\affiliation{Jinan Institute of Quantum Technology, Jinan 250101, China}

\author{Dian Wu}
\author{Bing Bai}
\author{Ming-Cheng Chen}
\author{Qi-Chao Sun}

\author{Jun Zhang}
\affiliation{Hefei National Research Center for Physical Sciences at the Microscale and School of Physical Sciences, University of Science and Technology of China, Hefei 230026, China}
\affiliation{Shanghai Research Center for Quantum Science and CAS Center for Excellence in Quantum Information and Quantum Physics, University of Science and Technology of China, Shanghai 201315, China}
\affiliation{Hefei National Laboratory, University of Science and Technology of China, Hefei 230088, China}

\author{Sixia Yu}
\affiliation{Hefei National Research Center for Physical Sciences at the Microscale and School of Physical Sciences, University of Science and Technology of China, Hefei 230026, China}

\author{Qiang Zhang}
\author{Chao-Yang Lu}
\author{Jian-Wei Pan}

\affiliation{Hefei National Research Center for Physical Sciences at the Microscale and School of Physical Sciences, University of Science and Technology of China, Hefei 230026, China}
\affiliation{Shanghai Research Center for Quantum Science and CAS Center for Excellence in Quantum Information and Quantum Physics, University of Science and Technology of China, Shanghai 201315, China}
\affiliation{Hefei National Laboratory, University of Science and Technology of China, Hefei 230088, China}
 
\begin{abstract}

Nonlocality captures one of the counterintuitive features of nature that defies classical intuition. Recent investigations reveal that our physical world's nonlocality is at least tripartite; i.e., genuinely tripartite nonlocal correlations in nature cannot be reproduced by any causal theory involving bipartite nonclassical resources and unlimited shared randomness. Here, by allowing the fair sampling assumption and postselection, we experimentally demonstrate such genuine tripartite nonlocality in a network under strict locality constraints that are ensured by spacelike separating all relevant events and employing fast quantum random number generators and high-speed polarization measurements. In particular, for a photonic quantum triangular network we observe a locality-loophole-free violation of the Bell-type inequality by 7.57 standard deviations for a postselected tripartite Greenberger-Horne-Zeilinger state of fidelity $(93.13 \pm 0.24)\%$, which convincingly disproves the possibility of simulating genuine tripartite nonlocality by bipartite nonlocal resources with globally shared randomness.
 
\end{abstract}

\date{\today}
\maketitle

Quantum theory allows correlations between spatially separated systems that are incompatible with local realism \cite{bell1964einstein}. The most well-known manifestation is the correlation in bipartite systems -- Bell nonlocality \cite{bell1964einstein, brunner2014bell} that originally lies in the nature of quantum entanglement. As confirmed via loophole-free violations of Bell inequalities \cite{hensen2015loophole, shalm2015strong, giustina2015significant, rosenfeld2017event, li2018test}, Bell nonlocality has found novel applications in many quantum information tasks such as device-independent quantum cryptography \cite{ekert1992quantum, acin2007device} and randomness certification \cite{acin2016certified, pironio2010random}.

In contrast to bipartite systems, multipartite systems display much richer and complex correlation structures \cite{pan2012multiphoton, brunner2014bell}. Histrionically, multipartite entanglement conventionally understood as the property of non-separability \cite{horodecki2009quantum} was used to violate Bell-like inequalities (e.g., Mermin's inequality \cite{mermin1990extreme}) for multipartite nonlocality \cite{pan2000experimental,erven2014experimental}. However, one could reproduce such Bell-like violations by using entanglement of partial separability. This fact was first point out by Svetlichny in 1987 \cite{svetlichny1987distinguishing}, who derived an inequality such that it is obeyed by three-particle bi-separable states but its violation shows the states are truly three-particle nonseparable. This motivates great interests in the study of the strongest form of multipartite nonlocality -- genuine multipartite nonlocality (GMN). 

\begin{figure}[!t]
	\centering
	\includegraphics[width=0.5\textwidth]{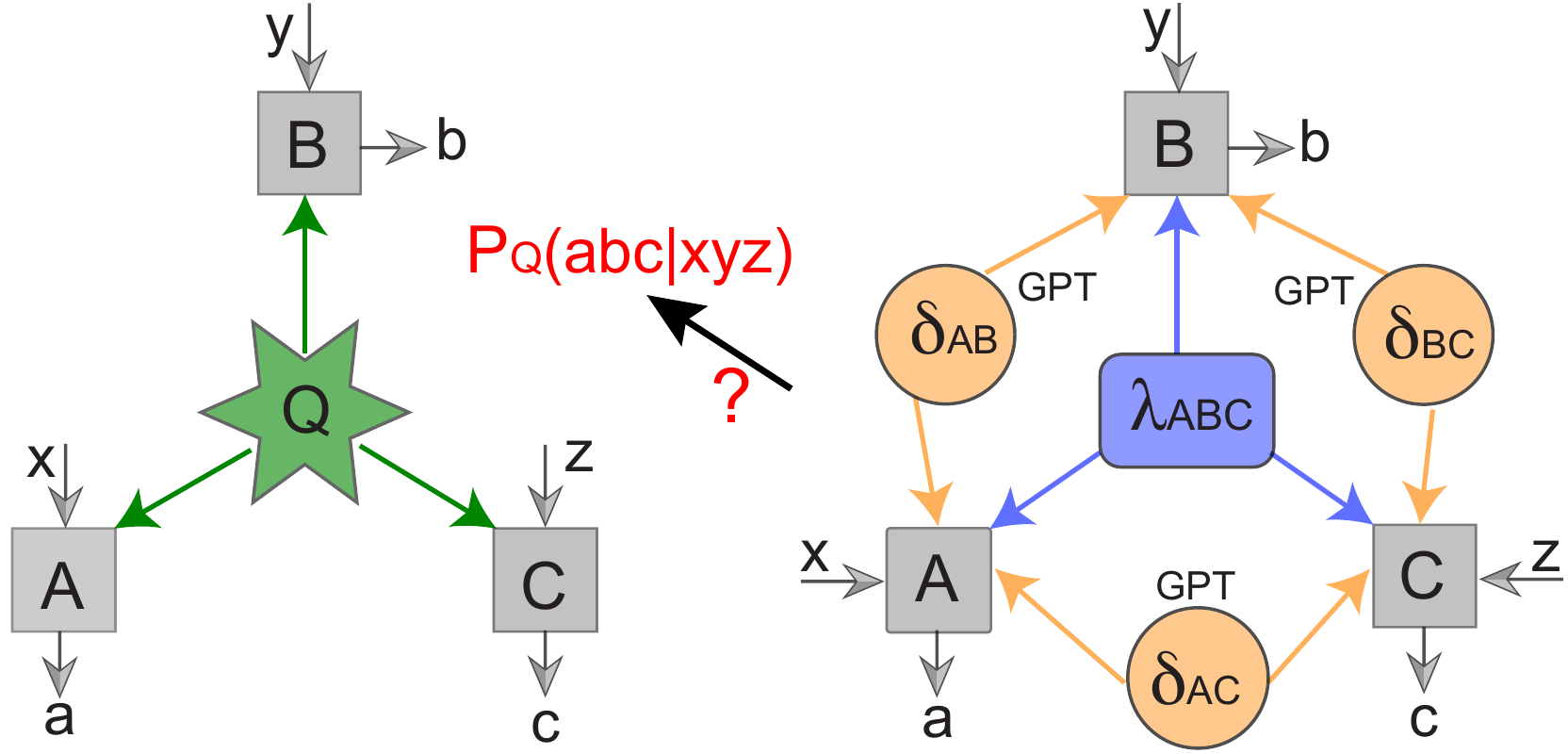}
	\caption{A triangular network features three observers (gray squares) A, B, and C for Alice, Bob, and Charlie, respectively, with ($x$, $a$), ($y$, $b$), and ($z$, $c$) being their inputs and outputs. The question of interest is whether or not the correlations $P_Q(abc|xyz)$ observed on the network on the left-hand side, in which each observer receives a particle from the tripartite-entangled quantum source (green starburst) and performs local measurements, can be simulated by the correlations obtained on the network on the right-hand side, in which the observers are connected by nonclassical bipartite resources ($\delta_{ij}$ with $i,j=A,B,C$) and shared randomness $ \lambda_{ABC}$.} 
	\label{Fig:concepts}
\end{figure}

In an effort to contribute to this line of research, Svetlichny's genuine tripartite nonlocality has been experimentally verified \cite{lavoie2009experimental, hamel2014direct} and generalized to scenarios featuring an arbitrary number of particles \cite{collins2002bell, seevinck2002bell} as well as arbitrary dimensions \cite{bancal2011detecting, chen2011detecting}. However, Svetlichny's GMN is relative to local operations and classical communication (LOCC) \cite{chitambar2014everything, gallego2012operational}. This is inconsistent with the situation involving space-like separated parties that enforces no-signaling condition \cite{gallego2012operational, bancal2013definitions}, which has already been shown in a table-top experiment \cite{zhang2016experimental}. Notably, restricted by non-signaling conditions, Svetlichny's GMN can also be observed in any network built by sharing only bipartite nonlocal resources, e.g., bipartite entanglement~\cite{contreras2021genuine}. Moreover, some correlations that display the forms of genuine tripartite nonlocality \cite{svetlichny1987distinguishing, bancal2013definitions} can be replicated by bipartite systems \cite{barrett2005nonlocal}. Realistically, all parties can have access to a common source of shared randomness. Also, instead of quantum theory, one could exploit alternative physical theories such as Box-world \cite{janotta2012generalizations} for nonclassical resources (e.g., Popescu-Rohrlich boxes \cite{popescu1994quantum}). Interestingly, it has been shown that Box-world theory cannot reproduce all quantum correlations even we allow globally shared classical randomness \cite{chao2017test, bierhorst2021ruling}. Furthermore, bipartite resources are not enough to reproduce tripartite phenomena in a theory-independent analysis, however, shared randomness is not involved in the analysis \cite{henson2014theory}. Thus, it is interesting to study GMN relative to local operations and shared randomness (LOSR) \cite{schmid2020understanding} from a theory-agnostic perspective, i.e., whether there are correlations in Nature irreproducible by sharing only fewer-partite nonlocal resources with LOSR (Fig. \ref{Fig:concepts}).

Recently, Coiteux-Roy \textit{et al.} answered positively by taking into account all causal theories compatible with device replication (i.e., refer to generalized probabilistic theories (GPTs)), including classical theory, quantum theory, non-signaling boxes, and any hypothetical causal theory \cite{coiteux2021no,coiteux2021any}. In the framework of LOSR, they refined Svetlinchny's GMN to genuine LOSR multipartite nonlocality or GMN in network. With the inflation technique widely used in analyzing theory-independent correlations \cite{wolfe2019inflation, wolfe2021quantum}, they derived a device-independent Bell-type inequality that is satisfied by all multipartite correlations arising from sharing fewer-partite nonlocal resources and global randomness. From the violations to the Bell-type inequality by $N$-partite Greenberger-Horne-Zeilinger (GHZ) states for all finite $N$, they thus proved that  Nature's nonlocality must be boundlessly multipartite in any causal GPTs.

In this Letter, we aim to show genuine LOSR tripartite nonlocality in a state-of-art photonic quantum network under strict locality constraints, i.e., all the parties involved be space-like separated. This requirement is crucial in analyzing Bell-type inequality violation as potential locality loopholes might be exploited by adversaries and also  enforces the non-signaling conditions with classical communication between the parties being forbidden. In details, we adopt post-selection and prepare a triggered three-photon GHZ state from two independent entangled pair sources, and distributed the state to three space-like separated observers Alice, Bob, and Charlie. The locality loophole is closed by space-like separating relevant events and using fast quantum random number generators (QRNGs) and high-speed polarization analyzers. We require the fair sampling assumption and post-selection in the experiment, and show that the produced tripartite correlations cannot be simulated by any bipartite nonlocal resources with LOSR, i.e., they are genuinely LOSR tripartite nonlocal. We expect our work will stimulate further experimental investigation of genuinely multipartite nonlocality to better understand our Nature.

The genuine LOSR tripartite nonlocality proposed by Coiteux-Roy \textit{et al.} is guaranteed by violations to the device-independent inequality arising from combining two intertwined games \cite{coiteux2021no,coiteux2021any}, respectively detecting (1) some nonclassical resources albeit possibly bipartite, and (2) some tripartite resource albeit possibly classical. For (1), the Bell game, they exploit standard Clauser-Horne-Shimony-Holt (CHSH) Bell test between Alice and Bob, conditioned on Charlie's output result $C_{1}=1$, which reads
\begin{equation} 
\begin{aligned} I _ {\text{ Bell} } ^ { C_{ 1 } = 1 } : = &\left\langle A _ { 0 } B _ { 0 } \right\rangle _ { C _ { 1 } = 1 } + \left\langle A _ { 0 } B _ { 1 } \right\rangle _ { C _ { 1 } = 1 } \\ &+ \left\langle A _ { 1 } B _ { 0 } \right\rangle _ { C _ { 1 } = 1 } - \left\langle A _ { 1 } B _ { 1 } \right\rangle _ { C _ { 1 } = 1 },
\label{Ibell}
\end{aligned} 
\end{equation} 
where subscript represents observer's setting choices and all observables take either $\pm 1$. In standard Bell game,  $I _ {\text{ Bell} } ^ { C_{ 1 } = 1 }$ can reach to $2\sqrt{2}$, which  necessitates nonclassical resources. For (2), all observers are required to give the same outputs, which can take either the two values $\pm1$. In this tripartite consistency game (i.e., Same game), the correlation is defined as \cite{coiteux2021no}
\begin{equation} 
	I _ { \text {Same} } : = \left\langle A _ {0} B _ {2} \right\rangle + \left\langle B _ {2} C _ {0} \right\rangle,
	\label{Isame}
\end{equation} 
and the perfect score is $I _ { \text {Same} }=2$.

We notice that $A_{0}:=A_{x=0}$ appears in both games, thus Alice cannot distinguish which of the two games she is participating in. This prevents her from playing the two games separately and she has to optimize Eq. \ref{Ibell} and \ref{Isame} simultaneously with her input $x=0$. Actually, it is impossible for Alice to decouple the two games, which indicates that performing well at both games (1) and (2) would require dependence on a genuinely LOSR tripartite nonlocal resources \cite{coiteux2021no}.

\begin{figure*}[!t]
	\centering
	\includegraphics[width=1\textwidth]{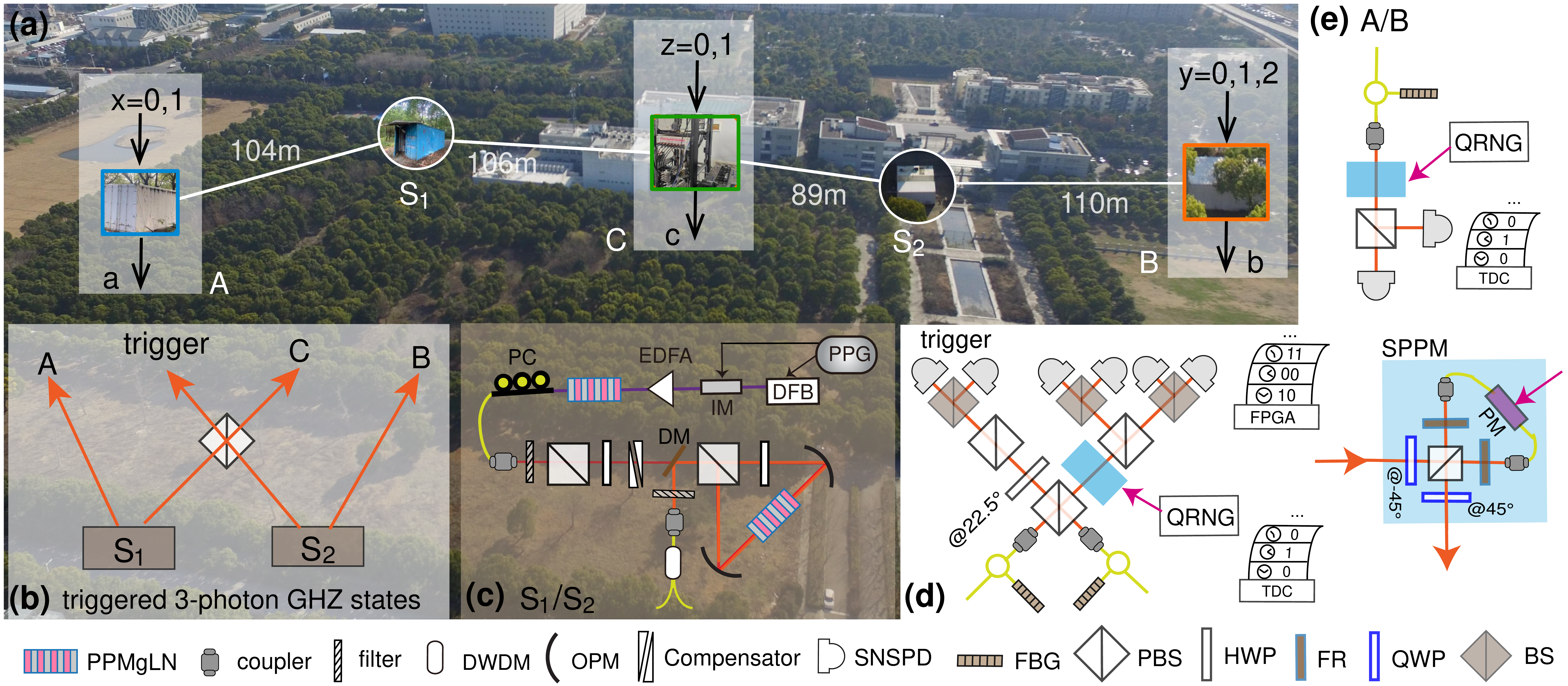}
	\caption{Scheme of the experiment. \textbf{(a)}, Three space-like separated observers (Alice, Bob, and Charlie) perform their local measurements. Two EPR sources $\text{S}_1$ and $\text{S}_2$ emit entangled photon pairs such that they are combined at a PBS for preparing a $\ket{\text{GHZ}_3}$ state. The beeline distances are 104 m, 106 m, 89 m, and 110 m between Alice-$\text{S}_{1}$, $\text{S}_{1}$-Charlie, Charlie-$\text{S}_{2}$, and $\text{S}_{2}$-Bob, respectively. Their optical fiber links are 112.63 m, 124.9 m, 109.6 m and 125.48 m respectively. \textbf{(b)}, The $\ket{\text{GHZ}_3}$ state is created by projecting one photon over the diagonal basis $\ket{+}$ (trigger) from a four-photon GHZ state after post-selection. \textbf{(c)}, In each source, a pair of polarization-entangled photons in state $\ket{\Phi^+}$ is prepared by pumping a PPMgLN crystal in a Sagnac-loop (details see text and Ref.\cite{sun2019experimental}). \textbf{(d)} and \textbf{(e)}, each observer performs local measurements on their received photon. The measurement choices are decided by their quantum random number generators (QRNGs). In each node, a high-speed single-photon polarization modulation (SPPM) is implemented to vary the direction of local polarization analysis. PPMgLMN periodically poled MgO doped lithium niobate; PC, polarization controller; DWDM, dense wavelength-division multiplexer; DM, dichroic mirror; OPM, off-axis parabolic mirror; FBG, fiber Bragg grating; BS, beam splitter; SNSPD, superconducting nanowire single photon detector.} 
	\label{Fig:diagrams}
\end{figure*}

With the inflation techniques \cite{wolfe2019inflation, wolfe2021quantum},  Coiteux-Roy \textit{et al.} then combine the two aforementioned games in one scenario. \textcolor{black}{If each two parties from three space-like separated observers Alice, Bob, and Charlie share a bipartite nonlocal resource and each party performs some local measurements, e.g.,  $A_{x}$, $B_{y}$, and $C_{z}$ with random inputs $x\in\{0,1\}$, $y\in\{0,1,2\}$, and $z\in\{0,1\}$ (Fig. \ref{Fig:diagrams} (a)), with outcomes $a,b,c=\pm1$, then the resulting  joint outcome probabilities $p(abc|xyz)$ satisfy the following device-independent Bell-type inequality (in slightly different but equivalent form)}
\begin{equation}
	F := I _ { \text {Bell } } ^ { C _ { 1 } = 1 } + \frac { 4 I _ { \text {Same } } - 8 } { 1 + \left\langle C _ { 1 } \right\rangle } \leqslant 2 \label{fdefination},
\end{equation}
where $F$ is the three-party correlation function and its calculations from $p(abc|xyz)$ are in Supplementary \cite{supp}.  A violation to the above inequality signatures the genuine LOSR tripartite nonlocality.

There are quantum correlations that violate the Bell-type inequality above. For example, we distribute tripartite GHZ state $\ket{\text{GHZ}_{3}}=(\ket{000}+\ket{111})/\sqrt{2}$ in a triangular network and set Alice's, Bob's, and Charlie's measurements as $A_{x}\in\{Z, X\}$, $B_{y}\in\{\frac{Z+X}{\sqrt{2}}, \frac{Z-X}{\sqrt{2}}, Z\}$, and $C_{z}\in\{Z, X\}$, respectively. Here $Z$ and $X$ are standard Pauil operators. In this case, the tripartite quantum strategy yields a maximum violation of $F=2\sqrt{2}$. For a mixture of the $\ket{\text{GHZ}_{3}}$ state with white noise, violation of the Eq. \ref{fdefination} requires a fidelity of $\geq93\%$. Note that Eq. \ref{fdefination} can be directly generalized to N-party GHZ state, however, the required state fidelity greatly increases with the system size N \cite{coiteux2021any} (details see \cite{supp}).

Our setup is shown in Fig. \ref{Fig:diagrams}. To violate Eq. \ref{fdefination}, we first prepare the $\ket{\text{GHZ}_{3}}$ state that can be efficiently created by combining two Einstein-Podolsky-Rosen (EPR) sources ($\text{S}_{1}$ and $\text{S}_{2}$) at a polarization beam splitter (PBS), as shown in Fig. \ref{Fig:diagrams} (b). We use a pulse pattern generator (PPG) to send out 250 MHz trigger signals, and the PPG in source $\text{S}_{2}$ acts as the master clock to synchronize all operations. In each source, a distributed feedback (DFB) laser is triggered to emit a 2 ns 1558 nm laser pulse, which is carved into 80 ps with an intensity modulator (IM). The laser pulses are frequency-doubled in a PPMgLN crystal after passing through an erbium-doped fiber amplifier (EDFA). We then use the produced 779 nm pump laser to drive a Type-0 spontaneous parametric down-conversion (SPDC) process in the second PPMgLN crystal in a polarization-based Sagnac loop. Each source produces pairs of photons in the Bell state $\ket{\Phi^{+}}=(\ket{HH}+\ket{VV})/\sqrt{2}$, where $H$ and $V$ denote horizontally and vertically polarization, respectively (see Fig. \ref{Fig:diagrams} (c) and \cite{sun2019experimental} for details). By interfering two photons at a PBS at Charlie's node, we get a four-photon GHZ state through the post-selection of four-fold coincidences, which is used for creating $\ket{\text{GHZ}_{3}}$ state when we measure the trigger photons in diagonal basis $\ket{+}$.

The observers then perform local measurements on their photon from the produced $\ket{\text{GHZ}_{3}}$ state. Alice and Charlie perform one of two measurements $A_{x}$ and $C_{z}$, respectively, while Bob measures one of three bases $B_{y}$. Their setting choices $x$, $z$, and $y$ are decided in real time by a fast quantum random number generator (QRNG) situated there. The QRNG at each station randomly outputs a sequence of bits with equal probabilities. Note that QRNG at Bob's station outputs four distinct sequences of two bits, however, we can discard one of the four outputs in order to only have three setting choices with equal probabilities for Bob under fair-sampling assumptions. All random bits from QRNG sources pass the NIST randomness tests \cite{rukhin2010statistical} (please refer to \cite{wu2022experimental} for more information about the QRNGs). To realize the fast measurement setting choice, we implemented a high-speed high-fidelity single-photon polarization modulation (SPPM), which consists of two fixed quarter-wave plates (QWP), two Faraday rotator (FR) and electro-optical phase modulator (PM), shown in Fig. \ref{Fig:diagrams} (see supplementary in Ref. \cite{wu2022experimental} for details). The SPPM varied photons' polarization at a rate of 250 MHz with a fidelity of \textcolor{black}{$\sim 99\%$} with random inputs. For Charlie, his setting choices decided by his QRNG were recorded with a time-to-digital converter (TDC). All his photon detection were analyzed in time and recorded by a field-programmable gate array (FPGA). Alice's and Bob's photon detection and setting results from QRNGs were recorded by their TDC, respectively. All the data were locally collected at the remote ports and sent to a separate computer to evaluate the three-party correlation function $F$.

\begin{figure}[!b]
	\centering
	\includegraphics[width=0.47\textwidth]{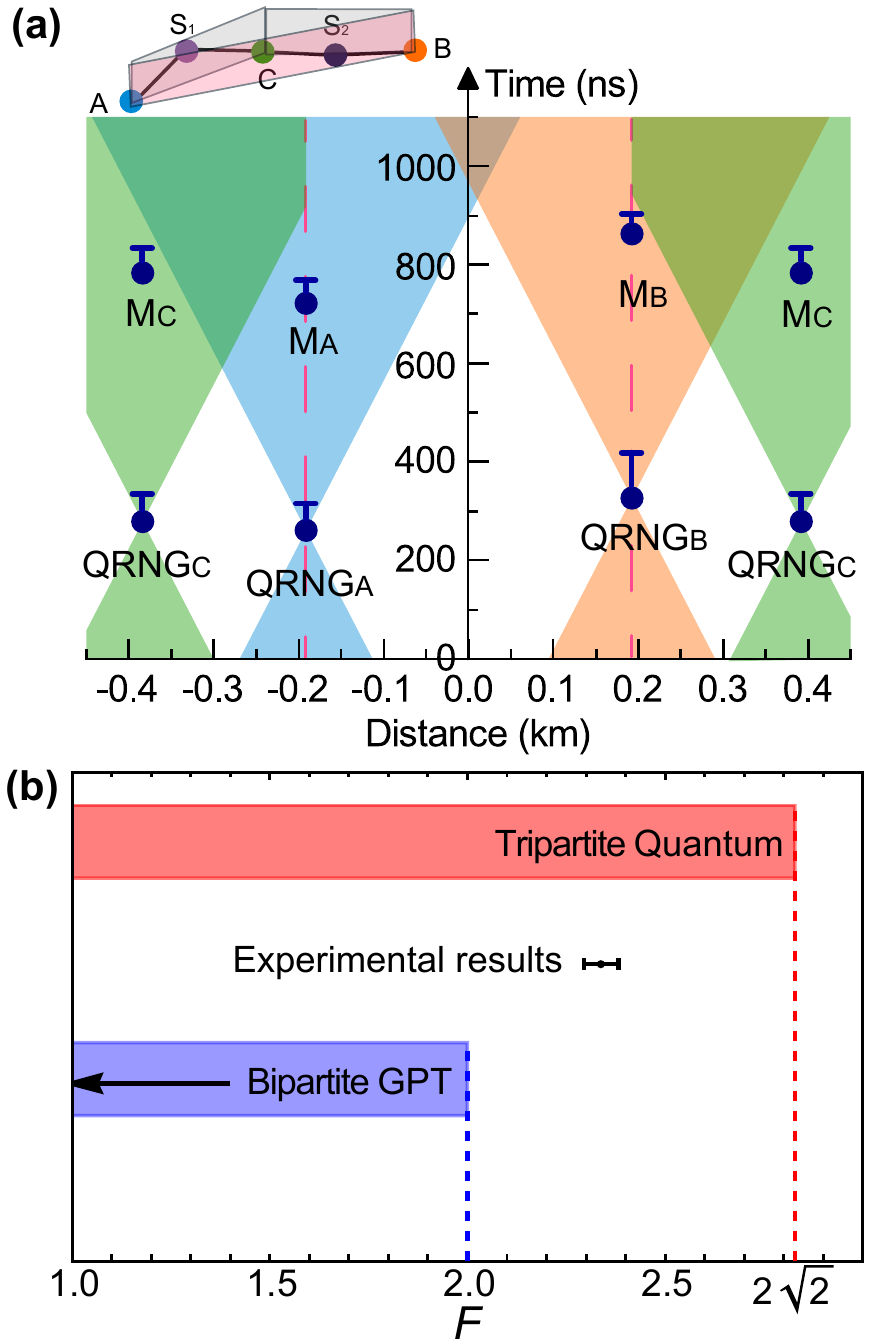}
	\caption{Space-time analysis and experimental results. \textbf{(a)} The left panel shows the space-like separations between the setting choice events by Charlie ($\text{QRNG}_{C}$) and Alice ($\text{QRNG}_{A}$), and between $\text{QRNG}_{C}$ ($\text{QRNG}_{A}$) and measurement events by Alice $\text{M}_{A}$ (Charlie, $\text{M}_{C}$), respectively. Similarly, the middle and right-hand region (split by red dashed lines) are for events events at Alice's and Bob's nodes, and events at Bob's and Charlie's nodes, respectively. Blue vertical bars denote the time elapsing for events, with start and end marked by circles and horizontal line, respectively. \textbf{(b)} The experimental result is 2.338(44), displaying as a black dot. Error bars indicate one standard deviations in the experiment.  Compared with the bounds of bipartite GPT (blue) and tripartite quantum (red), our result surpasses the bipartite GPT bound of 2 by 7.57 standard deviations, indicating genuinely LOSR tripartite nonlocality.} 
	\label{Fig:result}
\end{figure}

The timing and layout of the experiment are critical to close locality loopholes, such that for example any observer’s measurement results are causally independent from the others’ setting choices. Now considering Charlie and Alice, we space-like separate the events of Charlie completing the QRNG for setting choices ($\text{QRNG}_{C}$) from the events of finishing single photon detection by Alice ($\text{M}_{A}$), and vise versa. In each trial, the time elapses of a QRNG to generate a random bit that determine the setting choices for the received photon are both $53 \pm2$  ns for $\text{QRNG}_{C}$ and $\text{QRNG}_{A}$. The time elapse of measurement events is defined as the interval between a photon enters the loop interferometer in the SPPM (Fig. \ref{Fig:diagrams}) and the time of SNSPD outputs a signal for $\text{M}_{C}$ and $\text{M}_{A}$ are $44.9\pm0.5$ ns and $44.6\pm0.5$ ns, respectively. Their analysis are described in the left panel of Fig .\ref{Fig:result} (a), where $\text{M}_{A}$ is $156.3 \pm4$ ns outside the light cone of $\text{QRNG}_{C}$ and $\text{M}_{C}$ is $73.5\pm4$ ns outside the light cone of $\text{QRNG}_{A}$, satisfying the locality condition here. We summarize all relevant results for the other two slices in Fig. \ref{Fig:result} (a) (middle and right-hand panels), with the labels defined using the same conversion. All the time-space relations are drawn to scale. The analysis is summarized and detailed in \cite{supp}.

To estimate the fidelity of our prepared state after post-selection with respect to the ideal state $\ket{\text{GHZ}_{3}}$, we perform a fidelity witness that can be evaluated with only a few measurements. The average triggered three-photon rate is 0.3 Hz and the fidelity is calculated to be $93.13 \pm 0.24$ \%. We also perform a quantum state tomography to additionally characterize our prepared state (see \cite{supp}). We then evaluate the experimental violation of the inequality given by Eq. \ref{fdefination} and record 33770 four-fold coincidence detection events over 171725 seconds. As shown in Fig. \ref{Fig:result} (b), we obtain the correlation of $ F=2.338 \pm 0.044$, which is beyond the bipartite GPT bound by 7.57 standard deviations. That means the observed correlations via three-photon GHZ state cannot be reproduced by any two-way GPT resources with local operations and unlimited shared randomness, i.e., it is genuinely LOSR tripartite nonlocal \cite{coiteux2021no,coiteux2021any}.

Base on a optical quantum network under strict locality constraints, we have experimentally demonstrated that Nature's tripartite nonlocality cannot be simulated from any bipartite nonlocal causal theories. In our experiment, the locality loophole between the three parties is addressed by space-like separating relevant events and employing fast QRNGs and high-speed polarization analyzers. In this way, no information exchanges among the three parties in each trail, leading to the LOSR paradigm \cite{coiteux2021no,coiteux2021any}. Our demonstration requires fair-sampling assumptions and admits the post-selection loophole as well as the detection loopholes. To analyze Eq. \ref{fdefination}, we exclusively consider post-selection of the cases where the detectors click results in a unbiased sample under fair-sampling assumptions, which usually relates to detection loopholes \cite{larsson2014loopholes} that may be closed in the future with high-efficiency photon sources \cite{wang2019towards} and detectors. 

Another important issue is the post-selection for the entanglement generation process \cite{blasiak2021safe}, as our $\ket{\text{GHZ}_{3}}$ state depends on post-selecting a four-photon GHZ state, wherein one photon as a trigger co-located with Charlie's photon. With fair-sampling assumptions in each trial, we only consider performing post-selection on each port such that every party receives exactly one photon, although there are multiphoton events present that could decrease the prepared state fidelity. In the presence of post-selection, one could have selection bias that arises due to conditioning or restricting the data generated in the experiment \cite{blasiak2021safe}, which might lead to correlations breaking Bell-like inequality without necessarily claiming the genuine LOSR tripartite nonlocality. For example, if we use three independent bipartite entangled states shared by Alice, Bob and Charlie, and allow them measuring local parity operators, only post-selection on the interested events in the outcomes will lead to the statistics that show genuinely tripartite nonlocal features \cite{gebhart2021genuine,cao2022experimental}. However, one could potentially close the post-selection loophole at the sources by preparing states in a heralded event-ready manner such as using cascaded SPDC sources \cite{hamel2014direct, chaisson2021phase} or using on-demand single photon sources with fusion gates \cite{varnava2008good} in the future. Beyond the tripartite scenarios, a future interesting direction is to explore genuinely LOSR multipartite nonlocality in more complex networks, albeit it is experimentally challenging.

Note added -- After finishing our experiment, we became aware of two similar optical tabletop experimental works without closing locality loopholes \cite{cao2022experimental,mao2022test}.

\begin{acknowledgements}
$^{\ast}$These authors contributed equally to this work.

We thank Chang Liu and Quantum Ctek for providing the components used in the quantum random number generators. This work was supported by the National Natural Science Foundation of China, the Chinese Academy of Sciences, the National Fundamental Research Program, and the Anhui Initiative in Quantum Information Technologies.

\end{acknowledgements}

\bibliographystyle{unsrt}
\bibliography{refs}

\section{supplement information}

\subsection{N-party Inequality}
With the party N increasing, two intertwined three-player games $I_{\text{Bell}}^{C_{ 1 }= 1}$ and $I_{\text{Same} }$ will be extended as \cite{coiteux2021any}
\begin{equation*}
	\begin{aligned}
		I _ { \text {Bell } } ^ { \tilde { C } _ { 1 } = 1 } : =& \left\langle A _ { 0 } B _ { 0 } \right\rangle _ { \tilde { C } _ { 1 } = 1 } + \left\langle A _ { 0 } B _ { 1 } \right\rangle _ { \tilde { C } _ { 1 } = 1 } \\ &+ \left\langle A _ { 1 } B _ { 0 } \right\rangle _ { \tilde { C } _ { 1 } = 1 } - \left\langle A _ { 1 } B _ { 1 } \right\rangle _ { \tilde { C } _ { 1 } = 1 } ,\\
		I _{ \text {Same }_{ N }} : =& \left\langle A _ { 0 } B _ { 2 } \right\rangle + \left\langle B _ { 2 } C _ { 0 [ 1 ] } \right\rangle \\ &+ \left\langle C _ { 0 [ 1 ] } C _ { 0 [ 2 ] } \right\rangle + [ \ldots ] + \left\langle C _ { 0 [ N - 3 ] } C _ { 0 [ N - 2 ] } \right\rangle .
	\end{aligned}
\end{equation*}
where ${\tilde{C}}_1:=C_{1[1]}C_{1[2]}C_{1[...]}C_{1[N-2]}$ is defined over the collective of Charlies (i.e., ${\tilde{C}}_1=1$ means all Charlie players have input 1 and an even number of them output -1). The inequality of Eq.3 in the main text is then generalized to N parties as the follows 
\begin{equation*}
	F : = I _ { \text {Bell } } ^ { \tilde { C } _ { 1 } = 1 } + \frac { 4 I _ { \text {Same } _ { N } } - 4 ( N - 1 ) } { 1 + \left\langle \tilde { C } _ { 1 } ^ { 1 } \right\rangle } \leq 2.
	\label{inequalityN}
\end{equation*}


For a N-party GHZ state with white noise 
\begin{equation*}
	\rho = p\ket{\text{GHZ}_N}\bra{\text{GHZ}_N}+ (1-p)\frac{I}{2^N}
\end{equation*}
under optimal measurements, we have $I_{Bell}^{\tilde{C}_1}=2\sqrt{2}p$, $I _{\text {Same}_ { N }}=(N-1)p$ and $\mean{\tilde{C}_1^{1}}=0$. With Alice and Bob respectively perform measurement settings $A_{x}\in\{Z,X\}$ and $B_{y}\in\{\frac{Z+X}{\sqrt{2}},\frac{Z-X}{\sqrt{2}},Z\}$ while all parties (i.e., Charlie) perform setting $C_z \in\{Z,X\}$, the threshold $p$ and required fidelity $f$ could be derived as 
\begin{equation*}
	\begin{aligned}
		p \geq &\frac{2N - 1 }{ 2N - 2 + \sqrt { 2 } }, \\
		f \geq &\frac{2N - 1 +(\sqrt{2}-1)/2^N}{ 2N - 2 + \sqrt { 2 } }
	\end{aligned}
\end{equation*}
where we can see that the fidelity threshold increases when the system size N increases.

\subsection{Polarization measurement and experiment result}
\begin{table}[!b]
	\caption{Setting Configurations for each observer, $\phi_i$ is controlled by their QRNGs.}
	\renewcommand{\arraystretch}{1.4}
	\begin{tabular}{|>{\centering}m{1.9em}|c|c|>{\centering}m{3em}|c|c|}
		\hline
		\multicolumn{6}{|c|}{Alice's setting configuration}       \\ \hline
		\multicolumn{1}{|c|}{$x$} & QWP            & $\phi_1$ & QWP           & $A_x$ & $F_{Ax}$  \\ \hline
		0                         & $@-45^{\circ}$ & 0        & $@45^{\circ}$ & $Z$   & $99.23\%$ \\ \hline
		1                         & $@-45^{\circ}$ & $\pi/2$  & $@45^{\circ}$ & $X$   & $99.10\%$ \\ \hline
		\multicolumn{6}{|c|}{Bob's setting configuration}       \\ \hline
		\multicolumn{1}{|c|}{$y$} & QWP            & $\phi_2$ & QWP           & $B_y$          & $F_{By}$  \\ \hline
		0                         & $@-45^{\circ}$ & $\pi/4$  & $@45^{\circ}$ & $Z+X/\sqrt{2}$ & $98.89\%$ \\ \hline
		1                         & $@-45^{\circ}$ & $-\pi/4$ & $@45^{\circ}$ & $Z-X/\sqrt{2}$ & $99.17\%$ \\ \hline
		2                         & $@-45^{\circ}$ & $0$      & $@45^{\circ}$ & $Z$            & $99.45\%$ \\ \hline
		\multicolumn{6}{|c|}{Charlie's setting configuration}       \\ \hline
		\multicolumn{1}{|c|}{$z$} & QWP            & $\phi_3$ & QWP           & $C_z$ & $F_{Cz}$  \\ \hline
		0                         & $@-45^{\circ}$ & 0        & $@45^{\circ}$ & $Z$   & $99.34\%$ \\ \hline
		1                         & $@-45^{\circ}$ & $\pi/2$  & $@45^{\circ}$ & $X$   & $99.17\%$ \\ \hline
	\end{tabular}
	\label{basis conf}
\end{table}
To determine the value of three-party correlation in our experiment, Alice and Charlie need to randomly chose their measurement settings $A_x$, $C_z$ among $\{Z,X\}$, and Bob needs to chose his setting $B_y$ among $\{\frac{Z+X}{\sqrt{2}},\frac{Z-X}{\sqrt{2}},Z\} $. All the measurement settings could be achieved with the SPPM device shown in Fig.2 in main text that consists of two QWPs sandwiched by an EOPM. The EOPM consists of a PBS, two Faraday Rotators(FR) and an electro-optic phase modulator: With the assistance of two FRs, two back-propagating components split on the PBS are coupled to the slow axis of the fiber phase modulator. In addition, the PM is positioned 20 cm from the loop center so that the two polarization components arrive at the PM with a time difference of 2 ns, allowing us to adjust the phase of a single component individually. The matrix of an EOPM and a SPPM can be written as $P\left(\phi\right)=\left(\begin{matrix}1&0\\0&e^{i\phi}\\\end{matrix}\right)$ and $M\left(\phi\right)=ie^{\frac{i\phi}{2}}\left(\begin{matrix}\cos{\frac{\phi}{2}}&\sin{\frac{\phi}{2}}\\-\sin{\frac{\phi}{2}}&\cos{\frac{\phi}{2}}\\\end{matrix}\right)$. Then we can calculate each basis's phase which is controlled by quantum random number generators (QRNGs). The fidelity of random basis choice is defined as $F_m=C_r/(C_r+C_w)$, where $C_r$ represents the photons recorded by the correct superconducting nanowire single photon detectors (SNSPDs) and $C_w$ the photons recorded by wrong SNSPDs, respectively. We show all the configurations and their related fidelity for each observer in Table. \ref{basis conf}. 
\begin{figure}[!b]
	\centering
	\includegraphics[width=1\linewidth]{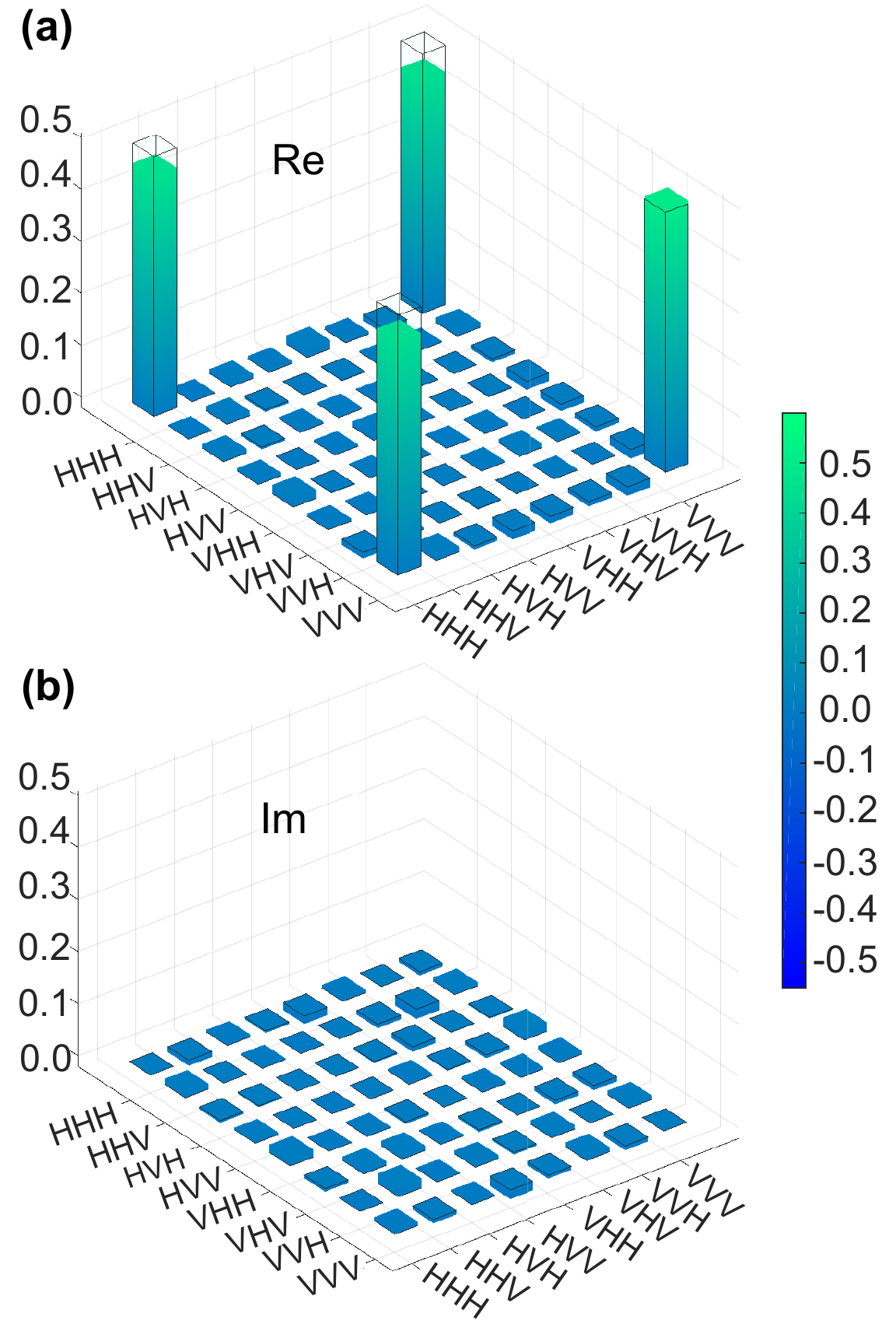}
	\caption{Tomography results of prepared $\ket{\text{GHZ}_{3}}$ state. \textbf{(a)} Real and \textbf{(b)} imaginary parts of the reconstructed density matrix. Black frame denotes the ideal state. The fidelity of $\ket{\text{GHZ}_{3}}$ state is ($93.22 \pm 0.25$)\%, where the standard deviation is estimated by the Monte Carlo analysis.}
	\label{fig:tomoghz}
\end{figure}


To firstly characterize our prepared $\ket{\text{GHZ}_3}$ state in experiment, we measure all over 27 settings $M_{ijk}=\sigma_i \sigma_j \sigma_j, (i,j,k \in \{1,2,3\})$ to perform state tomography, where $\sigma_i$ is the Pauli operator $\in \{X,Y,Z\})$. The tomography result is shown in Fig. \ref{fig:tomoghz}, which gives us a idea of the prepared state. However, tomographic approach could lead to biases that overestimate the entanglement \cite{schwemmer2015systematic}. Here we measure a fidelity witness to estimate the fidelity of our GHZ state with proper statistical error bars by using
\begin{equation*}
	F=\bra{\text{GHZ}_3} \rho \ket{\text{GHZ}_3}=tr(\rho\ket{\text{GHZ}_3}\bra{\text{GHZ}_3})
\end{equation*}
where 
\begin{equation*}
	\begin{aligned}
		\ket{\text{GHZ}_3}\bra{\text{GHZ}_3}=&\frac{1}{2}\left(\ket{HHH}\bra{HHH} +\ket{VVV}\bra{VVV} \right)\\
		&+\frac{1}{8}\left(XXX-YXY-XYX-YYX\right).
	\end{aligned}
\end{equation*}

Using the data collected over the 27 settings from the tomographic method, we calculate the state witness for the fidelity, giving $F=(93.13 \pm 0.24)\%$.

\begin{figure}[!h]
	\includegraphics[width=0.5\textwidth]{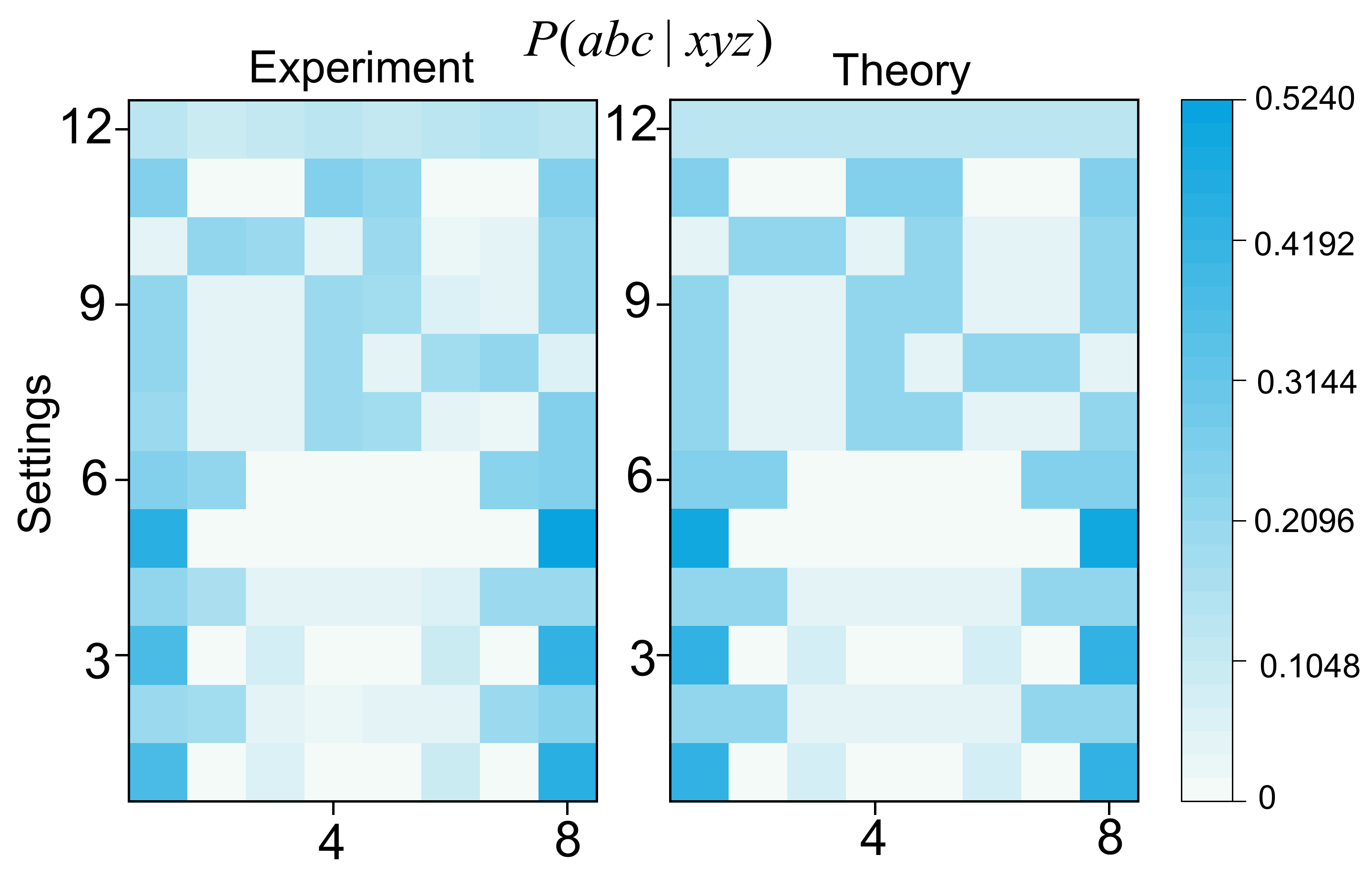}
	\caption{Result for the conditional joint probability distribution matrix $P(abc|xyz)$. Left panel shows the experimental results and right panel shows the theoretical prediction from an ideal experiment. The row index represents the measured strings $\{abc\}$ and the column index is the measurement setting $\{xyz\}$.}
	\label{Fig:Presults}
\end{figure}

For analysing the Bell-like inequality of Eq.3 in the main text, we totally collected 33770 four-fold coincidence events in 12 measurement setting combinations with 171725 seconds in the experiment. The result is shown in Table. \ref{measurement result}, from which we calculate the joint outcome probability $P(abc|xyz)$. The result of $P(abc|xyz)$ is shown in Fig. \ref{Fig:Presults}. 
\begin{table}[H]
	\centering
	\caption{Raw data recorded in the experiment. $xyz $ denotes basis choice and $abc $ denotes result string.}
	\renewcommand{\arraystretch}{1.5}
	\begin{tabular}{|c|c|c|c|c|c|c|c|c|}
		\hline
		\diagbox{$xyz$}{$abc$} & +++  & ++ - & +-+ & +- - & - ++ & - + - & - - + & - - -  \\ \hline
		000 & 1064 & 9   & 192 & 23  & 16  & 250 & 8   & 1227 \\ \hline
		001 & 590  & 538 & 105 & 92  & 133 & 126 & 570 & 688  \\ \hline
		010 & 993  & 17  & 193 & 21  & 14  & 263 & 6   & 1154 \\ \hline
		011 & 647  & 492 & 104 & 124 & 149 & 173 & 588 & 584  \\ \hline
		020 & 1225 & 8   & 9   & 15  & 18  & 24  & 5   & 1435 \\ \hline
		021 & 692  & 592 & 16  & 10  & 23  & 18  & 663 & 705  \\ \hline
		100 & 588  & 145 & 124 & 607 & 541 & 118 & 87  & 730  \\ \hline
		101 & 628  & 120 & 130 & 596 & 114 & 563 & 626 & 175  \\ \hline
		110 & 579  & 112 & 102 & 552 & 488 & 166 & 125 & 603  \\ \hline
		111 & 107  & 624 & 593 & 142 & 599 & 67  & 125 & 610  \\ \hline
		120 & 703  & 7   & 17  & 726 & 614 & 16  & 10  & 736  \\ \hline
		121 & 392  & 295 & 331 & 389 & 310 & 348 & 406 & 373  \\ \hline
	\end{tabular}
	\label{measurement result}
\end{table}

\begin{figure}[!t]
	\includegraphics[width=0.5\textwidth]{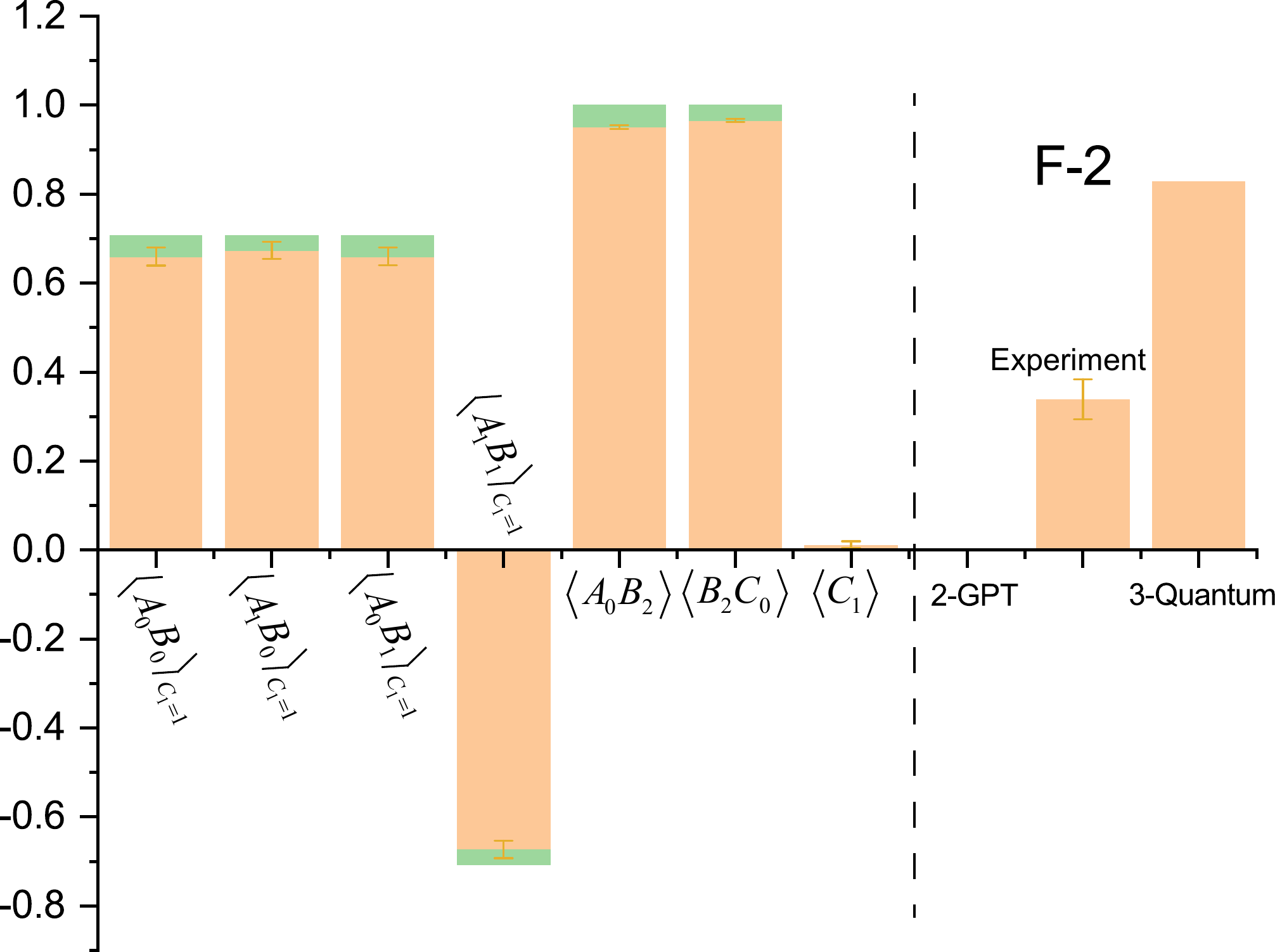}
	\caption{Experiment result. Relevant measurement result are shown in the left and violation of inequality $ F - 2 $ is shown in the right. Error bars represent one standard deviations in experiments.}
	\label{Fig:Fresults}
\end{figure} 
The three-party correlation function in the main text can be obtained from: 
\begin{equation*}
	\begin{aligned}
		\mean{A_x B_y}_{{C_1}=1} =& \frac{\sum_{a,b,c}^{z=1,c=+1}P(abc|xyz)ab}{\sum_{a,b,c}^{z=1,c=+1}P(abc|xyz)} \\
		\mean{A_x B_y} = & \sum_{a,b,c} P(abc|xyz)ab \\
		\mean{B_y C_z} = & \sum_{a,b,c} P(abc|xyz)bc \\
		\mean{C_z} = & \sum_{a,b,c} P(abc|xyz)c
	\end{aligned}
\end{equation*}

All the above results are shown in Fig \ref{Fig:Fresults}, together with the final violation of the inequality, bipartite GPT bound, and tripartite quantum bound (all the bounds are shown in $F-2$).

\subsection{Details about space-time diagram}
To determine the space time relation between three measurement stations, we measure the distances and fiber lengths  between all relevant nodes in our experiment. The space distance is measured with a laser rangefinder whose uncertainty is less than 0.4 cm, much smaller than the station size (on an optical table) of 1m. Therefore, we can put an upper bound of the distance uncertainty to be 1 m. After considering fiber fusion accuracy we set the uncertainty of fiber length to 0.1 m. All results are shown in Table. \ref{lengths}. 
\begin{table}[!b]
	\caption{Length for space and fiber links.}
	\renewcommand{\arraystretch}{1.5}
	\begin{tabular}{|c|c|c}
		\hline
		Link          & Space distance (m) & \multicolumn{1}{c|}{fiber length (m)} \\ \hline
		Alice-S1      & $104 \pm 1$        & \multicolumn{1}{c|}{$112.6 \pm 0.1$}  \\ \hline
		S1-Charlie    & $106 \pm 1$        & \multicolumn{1}{c|}{$124.9 \pm 0.1$}  \\ \hline
		Charlie-S2    & $89 \pm 1$         & \multicolumn{1}{c|}{$109.6 \pm 0.1$}  \\ \hline
		S2-Bob        & $110 \pm 1$        & \multicolumn{1}{c|}{$125.5 \pm 0.1$}  \\ \hline
		Alice-Charlie & $192 \pm 1$        &                                       \\ \cline{1-2}
		Alice-Bob     & $384 \pm 1$        &                                       \\ \cline{1-2}
		Bob-Charlie   & $199 \pm 1$        &                                       \\ \cline{1-2}
	\end{tabular}
	\label{lengths}
\end{table}

\begin{table}[!b]
	\caption{Time delays for space-time analysis.}
	\renewcommand{\arraystretch}{1.5}
	\begin{tabular}{|cc|}
		\hline
		\multicolumn{2}{|c|}{Alice's Photon Delays (ns)}     \\ \hline
		\multicolumn{1}{|c|}{Source delay} & $160.2 \pm 0.5$ \\ \cline{2-2} 
		\multicolumn{1}{|c|}{Fiber delay}  & $563.0 \pm 0.5$ \\ \cline{2-2} 
		\multicolumn{1}{|c|}{Measurement}  & $44.6 \pm 0.5$  \\ \cline{2-2} 
		\multicolumn{1}{|c|}{\textbf{Total}}        & $767.8 \pm 0.5$ \\ \hline
		\multicolumn{2}{|c|}{Bob's Photon Delays (ns)}       \\ \hline
		\multicolumn{1}{|c|}{Source delay} & $160.2 \pm 0.5$ \\ \cline{2-2} 
		\multicolumn{1}{|c|}{Fiber delay}  & $704.0 \pm 0.5$ \\ \cline{2-2} 
		\multicolumn{1}{|c|}{Measurement}  & $38.4 \pm 0.5$  \\ \cline{2-2} 
		\multicolumn{1}{|c|}{\textbf{Total}}        & $902.6 \pm 0.5$ \\ \hline
		\multicolumn{2}{|c|}{Charlie's Photon Delays (ns)}   \\ \hline
		\multicolumn{1}{|c|}{Source delay} & $160.2 \pm 0.5$ \\ \cline{2-2} 
		\multicolumn{1}{|c|}{Fiber delay}  & $624.5 \pm 0.5$ \\ \cline{2-2} 
		\multicolumn{1}{|c|}{Measurement}  & $44.9 \pm 0.5$  \\ \cline{2-2} 
		\multicolumn{1}{|c|}{\textbf{Total}}        & $829.6 \pm 0.5$ \\ \hline
	\end{tabular}
	\label{measure delays}
\end{table}

\begin{table}[!b]
	\caption{Time delays for measurement basis choice.}
	\renewcommand{\arraystretch}{1.5}
	\begin{tabular}{|cc|}
		\hline
		\multicolumn{2}{|c|}{Alice's Basis Selection Delays (ns)}   \\ \hline
		\multicolumn{1}{|c|}{QRNG delay}           & $53 \pm 2$     \\ \cline{2-2} 
		\multicolumn{1}{|c|}{Extra setting delay}  & $400 $         \\ \cline{2-2} 
		\multicolumn{1}{|c|}{Coaxial cable}        & $7.5 \pm 0.5$  \\ \cline{2-2} 
		\multicolumn{1}{|c|}{\textbf{Total}}                & $460.5 \pm 2$  \\ \hline
		\multicolumn{2}{|c|}{Bob's Basis Selection Delays (ns)}     \\ \hline
		\multicolumn{1}{|c|}{QRNG delay}           & $89 \pm 2$     \\ \cline{2-2} 
		\multicolumn{1}{|c|}{Extra setting delay}  & $444 $         \\ \cline{2-2} 
		\multicolumn{1}{|c|}{Coaxial cable}        & $1.5 \pm 0.5$  \\ \cline{2-2} 
		\multicolumn{1}{|c|}{\textbf{Total}}                & $534.5 \pm 2$  \\ \hline
		\multicolumn{2}{|c|}{Charlie's Basis Selection Delays (ns)} \\ \hline
		\multicolumn{1}{|c|}{QRNG delay}           & $53 \pm 2$     \\ \cline{2-2} 
		\multicolumn{1}{|c|}{Extra setting delay}  & $448 $         \\ \cline{2-2} 
		\multicolumn{1}{|c|}{Coaxial cable}        & $6.0 \pm 0.5$  \\ \cline{2-2} 
		\multicolumn{1}{|c|}{\textbf{Total}}                & $501 \pm 2$    \\ \hline
	\end{tabular}
	\label{basis delays}
\end{table}

\begin{table}[!h]
	\caption{Locality closures results}
	\renewcommand{\arraystretch}{1.5}
	\begin{tabular}{|cc|}
		\hline
		\multicolumn{2}{|c|}{Alice's Locality closure (ns)}   \\ \hline
		\multicolumn{1}{|c|}{From Bob}       & $842.8 \pm 4$  \\ \cline{2-2} 
		\multicolumn{1}{|c|}{From Charlie}   & $156.3 \pm 4$  \\ \hline
		\multicolumn{2}{|c|}{Bob's Locality closure (ns)}     \\ \hline
		\multicolumn{1}{|c|}{From Alice}     & $641.0 \pm 4$  \\ \cline{2-2} 
		\multicolumn{1}{|c|}{From Charlie}   & $44.9 \pm 4$   \\ \hline
		\multicolumn{2}{|c|}{Charlie's Locality closure (ns)} \\ \hline
		\multicolumn{1}{|c|}{From Alice}     & $73.5 \pm 4$  \\ \cline{2-2} 
		\multicolumn{1}{|c|}{From Bob}     & $163.9 \pm 4$   \\ \hline
	\end{tabular}
	\label{closure}
\end{table}
We set the moment when the pump light is emitted from $\text{S}_{1}$ and the moment when the entangled photons of Alice, Bob and Charlie are finally detected as the beginning and ending of one experimental trial. Table. \ref{measure delays} shows the delays for a photon arriving at each observer. We labeled the delay from the time $\text{S}_{1}$ generating a pump laser pulse to the time of entangled photons entering the fiber coupler as \textit{Source delay}. We then calculate the relative optical delay between the SPPM and output fiber coupler of $\text{S}_{1}$, defined as \textit{Fiber delay}. Finally, we measured the delay from the SPPM to the detector which is labelled as \textit{Measurement}.

The important times for basis choices are list in Table. \ref{basis delays}. The \textit{QRNG delay} is defined as the time elapse of quantum random number generation (from photodetector generating electronic signal to FPGA generating  random number data). The generated random data is delayed in FPGA before transmitting to the SPPM part, which is defined as the \textit{Extra setting delay}. The time from FPGA sending random number data to SPPM receiving the signal is defined as \textit{Coaxial cable} delay, including coaxial cable delay and short modulator driver delay.

With the data from Table. \ref{lengths}, Table. \ref{measure delays} and Table \ref{basis delays}, we can calculate the locality closures for each station, as shown in Table. \ref{closure}. For instance, the earliest basis choice for Bob can be calculated from his detection event:
\begin{equation*}
	\begin{aligned} 
		902.6 \mathrm {~ns} & \text { (Bob detection) } \\ 
		-38.4 \mathrm {~ns} & \text { (Measurement) } \\ 
		-534.5 \mathrm {~ns} & \text { (Bob's basis selection delay)} \\ 
		= 329.7 \mathrm {~ns} . & 
	\end{aligned}
\end{equation*}

The locality closure for Alice's detection and Bob's basis selection is:
\begin{equation*}
	\begin{aligned} 
		329.7 \mathrm {~ns} & \text { (Bob basis selection) } \\ 
		+\frac{384}{0.299792} \mathrm {~ns} & \text { (Travelling time for light) } \\ 
		-767.8 \mathrm {~ns} & \text { (Alice's detection) } \\ 
		= 842.8  \mathrm {~ns} . & 
	\end{aligned}
\end{equation*}

The upper uncertainty of space distance is set to $1$ m, with the speed of light being $ 0.299792$ m/ns, thus we can set the uncertainty of time closure as $4$ ns.

\end{document}